\documentclass[%
 reprint,
 amsmath,amssymb,
prl,
 longbibliography,
 lengthcheck,%
]{revtex4-1}

\usepackage{graphicx}
\usepackage{dcolumn}
\usepackage{bm}
\usepackage{hyperref}
\usepackage{txfonts}

\makeatletter

\newcommand{\Rmnum}[1]{\expandafter\@slowromancap\romannumeral #1@}
\makeatother

\begin{document}

\preprint{APS/123-QED}

\title{M\"ossbauer study of the field induced uniaxial anisotropy in electro-deposited FeCo alloy films}

\author{Zhiwei Li}\email{lizhiwei03@lzu.cn}
\author{Xu Yang}
\author{Haibo Wang}
\author{Xin Liu}
\author{Fashen Li}
 \affiliation{Institute of Applied Magnetics, Key Lab for Magnetism and Magnetic Materials of the Ministry of Education, Lanzhou University, Lanzhou 730000, Gansu, P.R. China.}


\date{\today}

\begin{abstract}
Thin ferromagnetic films with in-plane magnetic anisotropy are promising
materials for obtaining high microwave permeability. The paper
reports on the M\"ossbauer study of the field induced in-plane uniaxial
anisotropy in electro-deposited $FeCo$ alloy films.
The $FeCo$ alloy films have been prepared by electro-deposition method
with and without external magnetic field applied parallel to the
film plane during deposition. The vibrating sample magnetometry and
M\"ossbauer spectroscopy measurements at room temperature indicate
that the film deposited in external field
shows an in-plane uniaxial anisotropy with an easy direction
coincides with the external field direction and a hard direction
perpendicular to the field direction, whereas the film deposited
without external field doesn't show any in-plane anisotropy.
M\"ossbauer spectra taken in three geometric arrangements
show that the magnetic moments are almost
constrained in the film plane for the film deposited with
applied magnetic field. And the magnetic moments are
tend to align in the direction of the applied external
magnetic field during deposition, indicating that the
observed anisotropy should be attributed to directional ordering of atomic pairs.

\begin{description}
\item[PACC numbers]
3340,7570,7530G
\end{description}
\end{abstract}

\maketitle


\section{\label{sec:Intro}Introduction}
Ferromagnetic films have been investigated in a variety of systems due
to their wide range of applications \cite{1,2,3,4,5} and fundamental
theoretical interests \cite{6,7,8}. Applications such as inductors
, transformers and other magnetic devices require the magnetic
films with high cut-off frequency and high permeability \cite{1,3}.
And, the thin magnetic films with in-plane uniaxial anisotropy is essential
to achieve better high frequency properties in realizing the target of increasing
the static permeability and the resonance frequency simultaneously \cite{6}.
As is well known, an in-plane uniaxial anisotropy can be obtained with
oblique incidence of the vapor deposition beams \cite{9,10}, and
an easy direction of magnetization can also be
established parallel to the magnetic field applied during the
deposition \cite{11,12} or during magnetic annealing of the
specimens \cite{13,14}. The uniaxial anisotropy can also be
obtained by other means such as sputtering \cite{15},
electro-deposition \cite{16,17,18,19}, etc. Among these methods,
electro-deposition is an important cost-effective method for preparing
pure nanocrystalline metal and alloy films.

The characteristics of the induced magnetic anisotropy have been studied
comprehensively in the 1960s (for a review, see \cite{9,20,21} and
references therein). It has been suggested
that the anisotropy constant, $K_u$, is strongly dependent upon
the film composition and deposition conditions.
Different mechanisms have also been proposed mainly considering the anisotropy
of atomic pair ordering \cite{22,23}, the strain
anisotropy \cite{24} resulting from the constraint of the
magnetostriction strain imposed on the film by the substrate and the
shape anisotropy \cite{11,25}.
Recently, Jin Han-Min et al reported a combination model \cite{21} of
Pair-Strain-Shape anisotropy to explain major characteristics of
experiments in a broad range reasonably.
But none of these models can
explain all the experimental data and
describe the origin of the anisotropy satisfactorily.

For electro-deposited films, the anisotropy derived from directional
ordering of atomic pairs have been mostly considered \cite{16,18}.
However, direct experimental evidence for directional ordering of
atomic pairs is not yet reported, and prevents clear understanding
of the origin of the field induced uniaxial anisotropy.
In the present work, using electro-deposition method, we prepared $FeCo$ alloy
film with an in-plane uniaxial anisotropy by applying an external
magnetic field during deposition and, film deposited without
external magnetic field for comparison. The films are studied by M\"ossbauer
spectroscopy in three geometric arrangements in order to probe the
configuration of the magnetic moments in the film from a more
microscopic scale.

\section{\label{sec:Experiment}Experiments}
The $FeCo$ alloy films were electro-deposited from the bath containing
$FeSO_4\cdot7H_2O$ ($28\,g/l$), $CoSO_4\cdot7H_2O$ ($28\,g/l$),
$H_3BO_3$ ($25\,g/l$), and L-Ascorbic acid ($0.5\,g/l$) at the
current density of $20\,mA/cm^2$, and the temperature of the bath
was kept at room temperature. The magnetic field applied parallel to
the film plane during deposition is $1200\,Oe$.

The morphology and composition for both films were investigated
by scanning electron microscopy (HitachiS-4800, Japan) with energy dispersive
X-ray spectroscopy (SEM/EDS).
Magnetic properties of the films were measured by a vibrating sample
magnetometer (Lake Shore 7304, USA) (VSM). $^{57}Fe$ conversion electron M\"ossbauer
spectra (CEMS) of the samples were recorded by a conventional
M\"ossbauer spectrometer (Wissel, Germany) working in constant acceleration mode at
room temperature.
The $\gamma$-ray source of the M\"ossbauer spectrometer is 25\,mCi
$^{57}Co$ in Palladium matrix. $\alpha$-Fe was used to calibrate the
isomer shifts and the driver velocity.

\section{\label{sec:Results}Results and Discussion}

\begin{figure}[htp]
\includegraphics[width=8 cm]{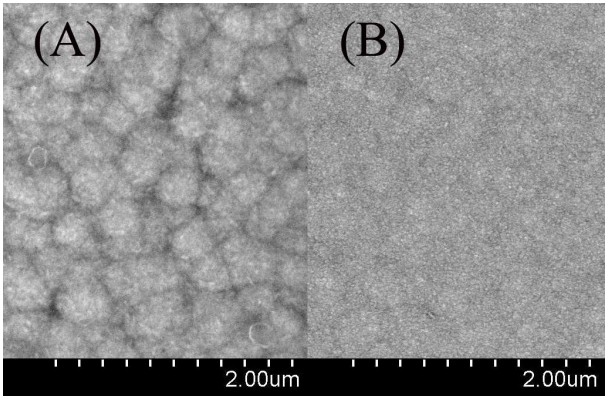}
\caption{\label{SEM} SEM images for film (A) deposited without external field and (B) deposited with external field..}
\end{figure}

Fig. \ref{SEM} shows the SEM images for film A deposited without the applied
external magnetic field and film B deposited with the external magnetic field.
It can be seen that the films are composed with small grains, and
the surface of film B is much smoother, which means that the grains
are smaller and composed much denser for film B than film A.
But, no macroscopic structure such as nonspherical grains
responsible for uniaxial anisotropy was observed.
This is consistent with the result reported by Fujita N et al \cite{18} 
in their electro-deposited $FeB$ films, and suggesting that the anisotropy should be attributed
to other origins of a more microscopic scale such as short-range
atomic pair ordering \cite{22}. Cross-sectional images obtained
from SEM revealed that the thickness for both film A and film B are about $1\,\mu m$.
EDS analysis shows that the composition is $Fe_{50}Co_{50}$ for both film
A and film B, and this is the very good composition to attain the largest
uniaxial magnetic anisotropy constant $K_u$ \cite{16} in electro-deposited $FeCo$ alloy films.

\begin{figure}[htp]
\includegraphics[width=8 cm]{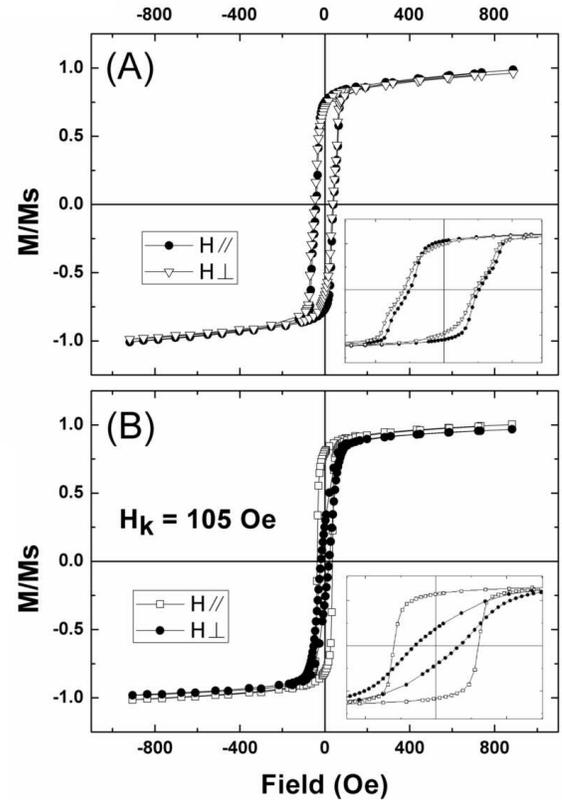}
\caption{\label{VSM} Typical in-plane hysteresis loops for film A and B at room temperature.
``$H_{\parallel}$'' and ``$H_{\perp}$'' denotes that the measurement
field is parallel and perpendicular to
the direction of the applied external magnetic field during deposition respectively.}
\end{figure}

Fig. \ref{VSM} shows the typical in-plane hysteresis loops
of the film deposited in the absence of external magnetic field (A)
and the film deposited in the presence of external magnetic field of $H_{ex}=1200\,Oe$ (B).
Here, ``$H_{\parallel}$'' and ``$H_{\perp}$'' denotes that the measurement
field is parallel and perpendicular to the direction of the applied
external magnetic field during deposition respectively.
It can be seen clearly from the inset figure in the right-lower corner
of (A) and (B) that the film deposited without external field exhibits
isotropy while a distinct uniaxial anisotropy is established in
the film deposited with applied external field, with its easy direction
being parallel to the applied field direction during deposition.
The anisotropy field, $H_k$,
obtained from the VSM result is about $105\,Oe$.

\begin{figure}[htp]
\includegraphics[width=8 cm]{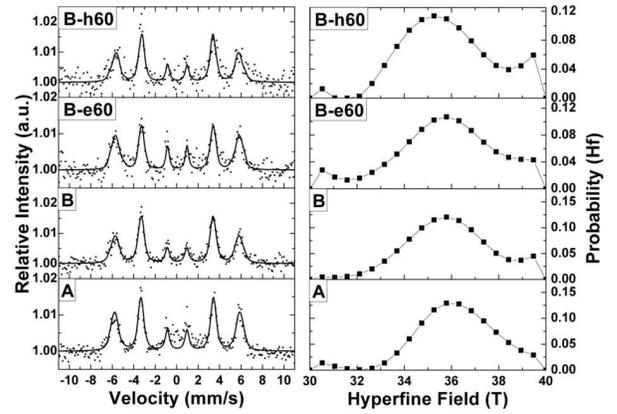}
\caption{\label{Moss} M\"ossbauer spectra taken at room temperature with the $\gamma$-rays
perpendicular to the film plane for (A) and (B), and with the $\gamma$-rays
forming an angle of $60\,^{\circ}$ with the easy direction (while
perpendicular to the hard direction) of film B for (B-e60) and
hard direction (while perpendicular to the easy direction) of film B for (B-h60).}
\end{figure}

In order to study the anisotropy deeply from a microscopic scale,
CEMS of the $FeCo$ alloy films were recorded at room temperature
in three geometric arrangements to probe the configuration of the
magnetic moments in the film. The CEMS of the alloy films are shown
in Fig. \ref{Moss}. As can be seen, the spectra are broad sextets because
different iron sites coexist due to random substitution of solid
solution \cite{26,27}, and therefore were fitted with the
hyperfine field distribution of a slightly modified version of the
Hesse and Rubartsch method using the Mosswinn programme \cite{28,29}.
Spectrum A and spectrum B were measured with the $\gamma$-ray beam perpendicular
to the film plane for film A and film B respectively.
Spectrum B-e60 and spectrum B-h60 were also measured for film
B, but with the $\gamma$-ray beam forming an angle of $60\,^{\circ}$
with the easy direction (while perpendicular to the hard direction) and
the hard direction (while perpendicular to the easy direction)
respectively. The fitted parameters are given in table \ref{TableMoss}. The hyperfine
field and isomer shift observed for both film A and film B are in agreement
with those reported by Sorescu M et al \cite{30} and Cohen N S et al \cite{26}.
But the hyperfine field of film B is slightly smaller than that of
film A, this may be caused by the differences of grain size and
density of the two films.

As is well known, using the M\"ossbauer effect, the
polar angle $\theta$ can be calculated from the relative intensities
of the hyperfine splitting of the M\"ossbauer lines. The relation
between $\theta$ and the relative intensities of $^{57}Fe$ lines are
given by \cite{31,32}
\begin{eqnarray}
I_{1,6} : I_{2,5} : I_{3,4} & & = 3(1+\cos^2\theta) : 4\sin^2\theta : (1+\cos^2\theta) \nonumber \\
& &=3 : x :1
\label{Ratio}
\end{eqnarray}
where ($0\leq x\leq 4$) corresponds to
($0\,^{\circ}\leq\theta\leq90\,^{\circ}$).
The polar angle $\theta$ calculated by equation
(\ref{Ratio}) from the value x in table \ref{TableMoss} are also
listed. The polar angle $\theta$ for film A($71.38\,^{\circ}$) is
much smaller than the value for film B($87.13\,^{\circ}$), which
could be attributed to the rough surface of film A,
 indicating that the applied magnetic field tends to align
the magnetic moments in the film plane.
If the magnetic moments were spatially isotropic
in the film plane, no differences of the polar angle should be
observed between B-e60 and B-h60. However, in our experiments the
$\theta$ obtained for B-e60($65.16\,^{\circ}$) is much smaller than
that for B-h60($80.32\,^{\circ}$), suggesting that the magnetic
moments of film B are anisotropic in the film plane and tend to align in the
direction of the easy direction of the film. This means that the
magnetic moments of film B have a directional ordering in the film
plane, and the ordering direction is parallel to the external magnetic
field applied during deposition.
de Oliveira L S et al \cite{27} have studied both external and internal
surface side of the electro-deposited films that detached from the substrate,
and no significant difference was observed.
So, we may neglect the effect of substrate on the formation of the
uniaxial anisotropy. K. Tanahashi and M.
Maeda \cite{16} have reported that the observed value of the
uniaxial anisotropy constant $K_u$ of their electro-deposited
$FeCo$ alloy films are considerably larger than that obtained for
the vacuum evaporated $FeCo$ alloy films, and by comparing the
theoretical dependence of $K_u$, the average internal stress
$\sigma$ and $(3/2)\lambda\sigma$ on the film composition with the
experimental data, they concluded that the observed anisotropy seems
to depend mainly on the atomic ordering.
And, as there was no macroscopic structure such as nonspherical grains giving rise to shape anisotropy
was observed from the SEM image, we suggest
that the observed anisotropy in our sample should also be attributed to
directional ordering of atomic
pairs \cite{22}, a microscopic phenomenon proposed by N\'{e}el.

As is shown above, our electro-deposited $FeCo$ alloy films have a distinct
in-plane uniaxial anisotropy. This is a perpetual requirement for magnetic films
in high-frequency applications \cite{6,15}. It is well known that for thin magnetic
films with in-plane anisotropy, the natural resonance frequency can be calculated by
the following equation \cite{33,34}
\begin{eqnarray}
f_r = \frac{\gamma}{2\pi}(4\pi M_sH_k)^{1/2}
\label{LLe}
\end{eqnarray}
where $\gamma$ is the gyro magnetic constant, $M_s$ is the
saturation magnetization and $H_k$ is the in-plane anisotropy
field. If we take the parameters $\gamma/2\pi = 2.8\, GHz/kOe$,
$4\pi M_s = 1.74\, T$ and use the anisotropic field  $H_k = 105\, Oe$,
we can obtain the resonance frequency of $3.8\, GHz$ which is
in good agreement with the measured data of $3.9\, GHz$. Detail experimental results
and model for the complex permeability of the film will be reported soon.

\begin{table}[ht]
\centering
\caption{M\"ossbauer parameters for $FeCo$ alloy films. $\delta$ is the isomer shift relative to $\alpha$-Fe,
$B_{hf}$ is the hyperfine field, $\Gamma_{exp}$ is the line width of the spectrum, x denotes the intensity
ratio of $I_{2,5}/I_{3,4}$. $\theta$ is the polar angle calculated from equation (\ref{Ratio}).}
\label{TableMoss} {\small
\begin{tabular}{c c c c c c c}
\hline \hline
Sample      &  $\delta$    &    $B_{hf}$      & $\Gamma_{exp}$ &  x    & $\theta$   &    \\
            &   (mm/s)     &    (T)           &   (mm/s)     &  (a.u.) & (Degree)   &       $\chi$          \\
\hline
A           &  0.047       &  36.167          &  0.439       &   3.26  &   71.38    &       1.84            \\
B           &  0.042       &  35.886          &  0.445       &   3.98  &   87.13    &       1.21            \\
B-e60       &  0.061       &  35.700          &  0.329       &   2.80  &   65.16    &       1.14            \\
B-h60       &  0.066       &  35.692          &  0.402       &   3.78  &   80.32    &       1.18            \\
\hline \hline
\end{tabular}}
\end{table}

\section{\label{sec:Conclusion}Concluding remarks}
In summary, $FeCo$ alloy film with a distinct in-plane uniaxial anisotropy
has been prepared by the simple electro-deposition method.
And the film is a good candidate for high frequency applications.
The uniaxial anisotropy was studied by means of
vibrating sample magnetometer and conversion electron M\"ossbauer
spectroscopy. VSM results show that the film deposited with an
magnetic field applied parallel to the film plane during deposition has an easy
direction coincides with that of the magnetic field, while the film
deposited without external magnetic field doesn't show any easy
direction in the film plane. M\"ossbauer spectra taken with the
$\gamma$-ray beam perpendicular to the film plane, show that the
magnetic moments are almost constrained in the plane for the film
deposited with external magnetic field. The spectra taken
with the $\gamma$-ray beam forming an angle of $60\,^{\circ}$ with the easy direction (while perpendicular to the
hard direction) and hard direction (while perpendicular to the easy direction) of the film give different
polar angles, indicating that the anisotropy of the film depend mainly on the
directional ordering of atomic pairs.

\begin{acknowledgments}
This work was supported by the National Natural Science Foundation of China under Grants No. 10774061.

\end{acknowledgments}


\end{document}